\documentclass[conference]{IEEEtran}
\ifCLASSINFOpdf
    \usepackage[pdftex]{graphicx}
\else
\fi
\usepackage{amsmath}
\usepackage{amssymb}
\usepackage{booktabs}
\usepackage{multirow}
\usepackage[table,xcdraw]{xcolor}
\usepackage{float}

\usepackage{algorithm}
\usepackage{algpseudocode}

\usepackage[hyphens]{url}
\usepackage[hidelinks]{hyperref}
\hypersetup{breaklinks=true}

\newcommand{\tableref}[1]{Table \ref{#1}}
\newcommand{\figureref}[1]{Fig. \ref{#1}}
\newcommand{\equationref}[1]{Equation (\ref{#1})}
\newcommand{\algorithmref}[1]{Algorithm \ref{#1}}
\newcommand{\sectionref}[1]{Section \ref{#1}}

\newcommand{\etal}{\textit{et al. }}

\newcommand{\bigO}[1]{\mathcal{O}({#1})}

\newcommand*\rot{\rotatebox{90}}

\begin{document}
%
\title{Community Detection for Power Systems Network Aggregation Considering Renewable Variability}

\author{
\IEEEauthorblockN{Raphael Araujo Sampaio, Gerson Couto Oliveira, Luiz Carlos da Costa Jr. and Joaquim Dias Garcia}
\IEEEauthorblockA{PSR, Rio de Janeiro, Brazil\\\{rsampaio, gerson, luizcarlos, joaquim\}@psr-inc.com}}


%



\maketitle

\begin{abstract}
The increasing penetration of variable renewable energy (VRE) has brought significant challenges for power systems planning and operation. These highly variable sources are typically distributed in the grid; therefore, a detailed representation of transmission bottlenecks is fundamental to approximate the impact of the transmission network on the dispatch with VRE resources. The fine grain temporal scale of short term and day-ahead dispatch, taking into account the network constraints, also mandatory for mid-term planning studies, combined with the high variability of the VRE has brought the need to represent these uncertainties in stochastic optimization models while taking into account the transmission system. These requirements impose a computational burden to solve the planning and operation models. We propose a methodology based on community detection to aggregate the network representation, capable of preserving the locational marginal price (LMP) differences in multiple VRE scenarios, and describe a real-world operational planning study. The optimal expected cost solution considering aggregated networks is compared with the full network representation. Both representations were embedded in an operation model relying on Stochastic Dual Dynamic Programming (SDDP) to deal with the random variables in a multi-stage problem.
\end{abstract}


%
\IEEEpeerreviewmaketitle

\section{Introduction}

The very fast insertion of variable renewable energy (VRE) sources such as wind and PV solar worldwide has brought significant economic and environmental benefits. However, the integration of large amounts of VRE capacity into the existing power systems has created new planning and operational challenges. For example, it is well known that 30 GW of wind power in Germany are backed up by almost the same capacity of thermal plants; the reason is that several days of zero wind have occurred.

The VRE integration challenges exist even in countries like Brazil, whose hydro plants (70\% of today’s 160 GW) with multi-year storage can serve as “water batteries” to manage VRE variability. The reason is that, in the next twenty years, the country’s power mix is likely to change significantly, with wind and solar PV dominating generation expansion and reaching $60$ GW and $55$ GW, respectively ($25$ GW utility scale and $30$ GW DG). This massive VRE share led to concerns about the disruption of regional power flow patterns and the increase of production uncertainty well beyond that of hydro generation.

\begin{figure}[H]
\centering
\includegraphics[width=3.0in]{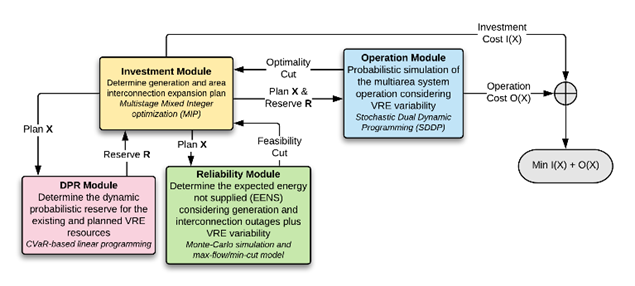}
\caption{Co-optimization of generation, reserve and area interconnections.}
\label{figure-main-components}
\end{figure}

These concerns motivated the development of new planning tools for the integrated planning (co-optimization) of generation, interconnection capacities and the generation reserve to handle the variability of existing and planned VRE sources. \figureref{figure-main-components} shows the main components of this new tool.

We observe in the \figureref{figure-main-components} that the optimal plan is obtained through the iterative solution of four modules: The investment module determines a trial expansion plan (represented by the vector $X$) and its associated investment cost, represented as $I(X)$ \cite{gorenstin1993power}. Next, the Dynamic Probabilistic Reserve (DPR) module calculates the generation reserve $R(X)$ required to handle the existing and planned VRE variability \cite{soares2019addressing}. The expansion plan and associated reserve are then sent to the operation module, which carries out a probabilistic simulation of the system operation and calculates the expected operation cost $O(X)$. Finally, the reliability module calculates the expected load curtailments due to generation and interconnection outages, combined with VRE variability.

Also, as seen in \figureref{figure-main-components}, the objective is to find a plan that minimizes the sum of investment and expected operation costs, subject to a supply reliability target (for example, the expected energy curtailment should not exceed $0.2\%$ of total yearly load). This optimality is assured by the Benders decomposition algorithm, in which the operation and supply reliability modules send linear constraints, known as Benders cuts, to the investment module, which is then re-solved, producing a new plan, and so on.

A key component of the above scheme and the focus of this paper is the aggregation of the transmission system, which in the case of Brazil has $10$ thousand buses and $14$ thousand circuits into a multi-area system. In addition to the significant computational benefits, this multi-area representation allows the planner to have a better understanding of the tradeoff between the costs of area generation reserve and of interconnections. The reason is that, in large countries such as Brazil, the spatial correlation of VRE production in different areas is usually small. Therefore, by investing in interconnections, we obtain a “portfolio effect” and require less reserve.      
Because of these conceptual and computational benefits, research interest in transmission network aggregation methods  has increased in the past years. In $2005$, Cheng \etal \cite{cheng2005ptdf} proposed a reduction method that approximates the power transfer distribution factors (PTDF) and the injection shift factors (ISF) of the original system. However, they are not capable of capturing the variability of congestion in scenarios. In $2013$, Cotilla-Sanchez \etal proposed a hybrid evolutionary algorithm, which uses the k-means \cite{lloyd1982least} as local improvement and a multiattribute objective function to partition the network, which takes account the electrical distances, the cluster sizes, the number of clusters, the intra-cluster cohesiveness, and intercluster connectedness. In $2016$, Stocker \etal \cite{stockertransmission} modified the \cite{cotilla2013multi}, by adding the LMP to the fitness function. In 2018, \etal \cite{cao2018incorporating} proposed a spectral clustering, also combined with the LMP. They analyzed a real-world application for network aggregation of Germany, comparing the total system costs obtained using the aggregated network with the reference expansion planning solution. In $2018$, the European Network of Transmission System Operators for Electricity (ENTSO-E) released a review \cite{entsoe2018} that proposed an algorithm that computes the differences of the LMPs of each circuit and starts grouping the buses with the smallest difference. This procedure is similar to find clusters using a minimum spanning tree, like Prim’s algorithm \cite{prim1957shortest}.

This paper presents a new aggregation methodology that uses the difference in nodal LMPs to identify transmission constraints; represents the variability of those LMPs; and complies with the requirement that the areas should be connected. The proposed approach is based on a “community detection technique” known as the Girvan–Newman Algorithm \cite{newman2004finding}. A capacitated graph represents the aggregated network. The number of nodes is chosen so as the mimic the major network bottlenecks using the locational marginal price (LMP) differences of buses as metric along the study horizon; generators and loads are mapped to their corresponding nodes.

The remainder of this article is structured as follows. \sectionref{section-methodology} presents the community detection methodology. \sectionref{section-casestudy} describes the case study, an operation planning study of Brazil, with the transmission bottlenecks observed, the network aggregation, and compare the total expected costs obtained with a simulation using the aggregated network. \sectionref{section-conclusion} draws conclusions and prospects for future work.
\section{Methodology}
\label{section-methodology}

Now we detail the community detection algorithm applied to power systems network aggregation. The algorithm can be better explained by abstracting some of power system's details and thinking of its network as a graph in which vertices are buses and edges are circuits, all the other parameters will be ignored. In the following paragraphs we will associate weights to the edges of the graph, however, these will not have straight relation to the physical parameters of the network.

\subsection{Community Detection}

A fundamental concept for community detection algorithms is the betweenness centrality. Betweenness centrality is a measure based on shortest paths - or minimum weight paths - introduced by Anthonisse in $1971$ \cite{anthonisse1971rush} and, later, generalized by Freeman in 1977 \cite{freeman1977set}. Given a graph $G = (V, E)$, where $V$ is the set of vertices and $E$ is the set of edges, let $\sigma_{st}$ be the number of shortest paths between the vertices $s \in V$ and $t \in V$ . Let $\sigma_{st}(v)$ be the number of shortest paths between $s$ and $t$ that includes the vertex $v \in V$. \equationref{eq-betweenness} defines the betweenness centrality of the vertex $v$.

\begin{align}
    g(v)= \sum_{s \neq v \neq t} \frac{\sigma_{st}(v)}{\sigma_{st}}
    \label{eq-betweenness}
\end{align}

The concept of betweenness centrality was also extended to edges, the edge betweenness centrality \cite{girvan2002community} is similarly defined by the number of shortest paths going through an edge. In a weighted graph, where each edge has a numerical value associated, weights are interpreted as distances: and this is one of the main reasons for choosing Girvan-Newman. Edges with smaller weights are more likely to belong to the same community. The worst-case time complexity to compute the edge betweenness of all vertices in a weighted graph is $\bigO{EV + V^2 \log V}$ and consumes a $\bigO{V+E}$ space \cite{brandes2001faster}.

Now we can detail the selected algorithms for community detection. 
The Girvan–Newman Algorithm \cite{newman2004finding} belongs to a hierarchical class of clustering methodologies \cite{fortunato2010community} and detects communities in graphs based on the edge betweenness measure. The algorithm progressively computes the edge betweenness centrality of the vertices and removes the highest score edge. After several steps, the graph starts to break into communities. When the iteration process is over, the algorithm produces a dendrogram, where the root is the full graph, and the leaves are the vertices. The worst-case time complexity of the algorithm is $\bigO{E^2V}$. \algorithmref{girvannewman-general} describes the general structure of Girvan–Newman:

\begin{algorithm}[H]
    \caption{Girvan–Newman}
    \begin{algorithmic}[1]
        \State Compute the betweenness centrality (score) of all edges
        \While{the graph is not empty}
            \State Remove the edge with highest score from the graph
            \State Recompute the score of all remaining edges
        \EndWhile
    \end{algorithmic}
    \label{girvannewman-general}
\end{algorithm}

\subsection{Proximity Measure - Weights on Edges}

The critical question now is how to define the edges' weights. We seek to identify boundaries of interchanges between communities representing eventual electric grid bottlenecks. We perform the identification by preserving the marginal cost of the network vertices within communities homogeneous; this means that no critical network constraints are active and, consequently, no significant bottlenecks are inside the community. Neighboring vertices with the same average marginal cost correspond to regions where bottlenecks do not occur due to internal transmission network constraints.

We define the distance between neighboring (connected) vertices as the absolute value of the difference between their respective nodal marginal costs. Given a sample of $n$ marginal cost of two neighboring vertices $v_1$ and $v_2$, defined as $\pi_{v_1}$ and $\pi_{v_2}$ respectively, \equationref{equation-distance} describes the distance between them.

\begin{align}
    \text{distance}_{v_1, v_2} = \frac{1}{n} \sum_{i = 1}^{n} \lvert \pi_{v_1}^{(i)} - \pi_{v_2}^{(i)}\rvert
    \label{equation-distance}
\end{align}

It is possible to assess LMPs by computing the optimal stochastic operating policy of a system considering the power grid. It is possible to portray the locational signals resulting from congestion in transmission bottlenecks, reflected in differences in nodal marginal costs.

\subsection{Complete Algorithm}

\algorithmref{algorithm-communitydetection} summarizes the three steps of the community detection for power systems: compute the optimal stochastic operating policy of a system, representing electricity grid, evaluate the distance for each vertex, and compute the Girvan–Newman Algorithm given the distances and topology of the network.

\begin{algorithm}[H]
    \caption{Community Detection for Power Systems}
    \begin{algorithmic}[1]
        \State Compute the optimal stochastic operating policy
        \For{each edge in the graph}
            \State Evaluate the distance of two neighboring vertices
        \EndFor
        \State Compute the Girvan–Newman Algorithm
    \end{algorithmic}
    \label{algorithm-communitydetection}
\end{algorithm}
\section{Case Study}
\label{section-casestudy}

Throughout this section, we analyze our algorithm applied to a Brazilian power system. First of all, we compute the optimal stochastic operating policy considering the electricity grid, using the SDDP algorithm \cite{pereira1991multi}.

The neighboring vertices with the same average marginal cost correspond to regions where bottlenecks do not occur due to internal transmission network constraints, meaning that the operating policy obtained by ignoring the transmission limits internal to each region would be similar to the policy obtained representing the full transmission network.

Once the algorithm returns the network aggregation, we define the interconnection limits between each pair of communities and represent the edges that interconnect them as circuits with flow limits equivalent to the sum of limits of the interconnecting circuits. In the aggregated model we also ignore the Kirchoff voltage law. In order to process the SDDP considering the proposed aggregation, we aggregate all generators and demands of each community in a single vertex representing the community.

\subsection{Brazilian Power System}
\

We evaluate our algorithm with a $2041$ Brazilian case to assess the quality of aggregate network modeling. Brazil’s network has $9645$ vertices and $14681$ edges in the high voltage network. Moreover we consider $152$ thermal plants, $1806$ renewable energy plants (solar, wind, biomass, small hydros) and $179$ hydro plants, $79$ of which have large reservoirs. As typical from the SDDP framework, the state-space has $79+179L$ states, where $L$ is the number of lags on the inflow stochastic process, $L=6$ in this case. The number of scenarios considered in this study is $200$ and the number of stages is $12$. We consider a given future cost function in the end of the $12^{th}$ stage. This is a fictitious approximation of the Brazilian power system in the year $2041$.

\subsection{Aggregated System}

Given a pre-processed case of the Brazilian power system, we can use the spot prices and apply the community detection algorithm, setting the number of communities to $7$. Because the underlying graph is large, the community detection step is moderately large, taking $31$ minutes. The obtained communities and their main characteristics are presented in the sequence.

\begin{figure}[!ht]
\centering
\includegraphics[width=3.0in]{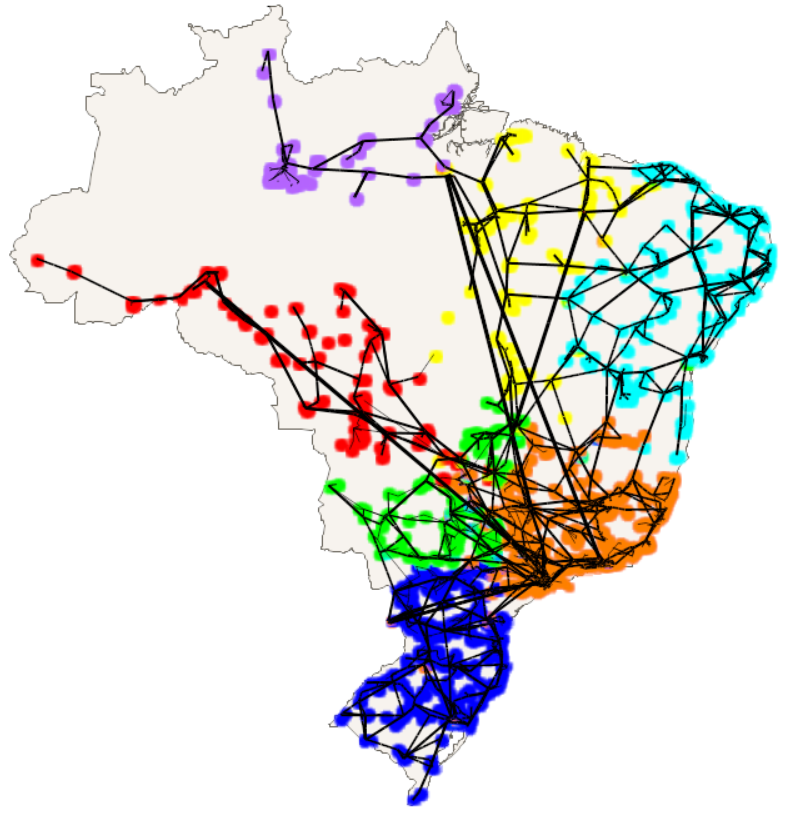}
\caption{Seven communities of the Brazilian study case.}
\label{figure-colombia-aggregation}
\end{figure}

\figureref{figure-colombia-aggregation} shows the network buses above $138$ kV as vertices, high voltage circuits as edges, and the seven communities obtained with our algorithm.

\tableref{table-colombia-analysis} shows the following characteristics for each community:
\begin{enumerate}
    \item The representative color in \figureref{figure-colombia-aggregation};
    \item The number of vertices;
    \item How many edges are internal to the community;
    \item How many edges connect it to neighboring communities;
    \item The average marginal cost of the vertices (\$/MWh);
    \item The average distance of the edges connecting their vertices (internal edges);
    \item The average distance of the edges connecting the community to another neighboring community (boundary edges).
\end{enumerate}

\begin{table}[H]
\centering
\caption{Analysis of each community.}
\label{table-colombia-analysis}
\begin{tabular}{@{}cccccccc@{}}
\toprule
\multirow{7}{*}{\rot{Community}} & \multirow{7}{*}{\rot{Color}}        & \multirow{7}{*}{\rot{\# of Vertices}} & \multirow{7}{*}{\rot{\# of Edges}} & \multirow{7}{*}{\rot{\# of Border Edges}} & \multirow{7}{*}{\rot{\begin{tabular}[c]{@{}c@{}}Marginal Cost \\ (\$/MWh)\end{tabular}}} & \multirow{7}{*}{\rot{Edges Weights}} & \multirow{7}{*}{\rot{\begin{tabular}[c]{@{}c@{}}Border Edges \\ Weights\end{tabular}}} \\
                                 &                                     &                                      &                                   &                                          &                                                                                         &                                     &                                                                                       \\
                                 &                                     &                                      &                                   &                                          &                                                                                         &                                     &                                                                                       \\
                                 &                                     &                                      &                                   &                                          &                                                                                         &                                     &                                                                                       \\
                                 &                                     &                                      &                                   &                                          &                                                                                         &                                     &                                                                                       \\
                                 &                                     &                                      &                                   &                                          &                                                                                         &                                     &                                                                                       \\
                                 &                                     &                                      &                                   &                                          &                                                                                         &                                     &                                                                                       \\ \midrule
1                                & \color[HTML]{fe7f00} $\blacksquare$ & \multicolumn{1}{r}{3097}             & \multicolumn{1}{r}{3783}          & \multicolumn{1}{r}{49}                   & \multicolumn{1}{r}{146.80}                                                              & \multicolumn{1}{r}{1.0}             & \multicolumn{1}{r}{5.7}                                                         \\
2                                & \color[HTML]{0101fe} $\blacksquare$ & \multicolumn{1}{r}{2267}             & \multicolumn{1}{r}{2848}          & \multicolumn{1}{r}{12}                   & \multicolumn{1}{r}{152.3}                                                               & \multicolumn{1}{r}{0.6}             & \multicolumn{1}{r}{4.9}                                                         \\
3                                & \color[HTML]{01fefe} $\blacksquare$ & \multicolumn{1}{r}{1548}             & \multicolumn{1}{r}{1669}          & \multicolumn{1}{r}{17}                   & \multicolumn{1}{r}{71.8}                                                                & \multicolumn{1}{r}{0.6}             & \multicolumn{1}{r}{8.7}                                                         \\
4                                & \color[HTML]{01fe01} $\blacksquare$ & \multicolumn{1}{r}{975}              & \multicolumn{1}{r}{1180}          & \multicolumn{1}{r}{41}                   & \multicolumn{1}{r}{153.8}                                                               & \multicolumn{1}{r}{0.7}             & \multicolumn{1}{r}{5.5}                                                         \\
5                                & \color[HTML]{fefe01} $\blacksquare$ & \multicolumn{1}{r}{816}              & \multicolumn{1}{r}{948}           & \multicolumn{1}{r}{26}                   & \multicolumn{1}{r}{114.9}                                                               & \multicolumn{1}{r}{2.0}             & \multicolumn{1}{r}{8.0}                                                          \\
6                                & \color[HTML]{fe0101} $\blacksquare$ & \multicolumn{1}{r}{637}              & \multicolumn{1}{r}{729}           & \multicolumn{1}{r}{15}                   & \multicolumn{1}{r}{152.8}                                                               & \multicolumn{1}{r}{1.2}             & \multicolumn{1}{r}{2.8}                                                         \\
7                                & \color[HTML]{b364fd} $\blacksquare$ & \multicolumn{1}{r}{286}              & \multicolumn{1}{r}{355}           & \multicolumn{1}{r}{2}                    & \multicolumn{1}{r}{162.3}                                                               & \multicolumn{1}{r}{0.7}             & \multicolumn{1}{r}{0.1}                                                         \\ \bottomrule
\end{tabular}                                                                                                                                                               
\end{table}

We highlight the average distance of the border circuits, which is greater than the average distance of the internal circuits of each region, except region four, with values close to each other. We also highlight that there are relatively few border circuits between regions.

\tableref{table-plants} shows the total number of plants of each type and allocation of each plant type in the communities.

\begin{table}[H]
\centering
\caption{Number of plants in each community.}
\label{table-plants}
\begin{tabular}{@{}lrrrrrrrr@{}}
\toprule
\multicolumn{1}{c}{Communities} & \multicolumn{1}{c}{$1$} & \multicolumn{1}{c}{$2$} & \multicolumn{1}{c}{$3$} & \multicolumn{1}{c}{$4$} & \multicolumn{1}{c}{$5$} & \multicolumn{1}{c}{$6$} & \multicolumn{1}{c}{$7$}& \multicolumn{1}{c}{Total} \\ \midrule
Hydro Plants                    & $74 $                   & $47 $                   & $ 6 $                   & $16 $                   & $11$                    & $ 19$                   & $ 6$                    & $179$                      \\
Thermal Plants                  & $45 $                   & $27 $                   & $ 36$                   & $10 $                   & $14$                    & $ 4 $                   & $16$                    & $152$                     \\
Renewable Plants                & $413$                   & $298$                   & $721$                   & $188$                   & $88$                    & $ 89$                   & $ 9$                    & $1806$                     \\ \bottomrule
\end{tabular}
\end{table}

\tableref{table-capacity} presents the installed capacity of each plant type in each community.

\begin{table}[H]
\centering
\caption{Installed capacity of each community, in GW.}
\label{table-capacity}
\begin{tabular}{@{}lrrrrrrr@{}}
\toprule
\multicolumn{1}{c}{Communities} & \multicolumn{1}{c}{$1$} & \multicolumn{1}{c}{$2$} & \multicolumn{1}{c}{$3$} & \multicolumn{1}{c}{$4$} & \multicolumn{1}{c}{$5$} & \multicolumn{1}{c}{$6$} & \multicolumn{1}{c}{$7$}\\ \midrule
Hydro Plants                    & $37.2$ & $22.2$ & $9.1$ & $8.3$ & $12.6$ & $11.9$ & $1.2$                                    \\
Thermal Plants                  & $26.1$ & $6.8$ & $5.4$ & $0.3$ & $1.9$ & $0.5$ & $0.9$                                   \\
Renewable Plants                & $28.6$ & $22.3$ & $75.8$ & $13.4$ & $8.1$ & $4.8$ & $0.7$                                    \\ \bottomrule
\end{tabular}
\end{table}

\subsection{Comparing Representations}

\tableref{table-operativecost} compares the operative cost of the two network representations. Some statistics are included to characterize the operation on all the scenarios. Note that the percentage difference between the averages is less than $0.4\%$, which is a first hint that the approximation is reasonable and might be a good approximation.

\begin{table}[H]
\centering
\caption{Operative cost, in $10^6 \$$.}
\begin{tabular}{@{}lcccc@{}}
\toprule
\multicolumn{1}{c}{Study} & Average & Standard Deviation   & Minimum   & Maximum   \\ \midrule
Original                  & $17522$   & $5280$ & $14027$ & $52221$ \\
Aggregated                & $17591$   & $4873$ & $13991$ & $44496$ \\ \bottomrule
\end{tabular}
\label{table-operativecost}
\end{table}

Table \ref{table-time} present the time required to simulate all $200$ scenarios of the two cases on a single threaded \textit{Intel Core i7 8th Generation 1.8 GHz}. We can see that the time reduction was larger than $10$ times.

\begin{table}[H]
\centering
\caption{Computational cost.}
\begin{tabular}{@{}lc@{}}
\toprule
\multicolumn{1}{c}{Study} & Execution Time (s)   \\ \midrule
Original                  & $4166$   \\
Aggregated                & $384$   \\ \bottomrule
\end{tabular}
\label{table-time}
\end{table}

Now we present some detailed results in order to show that aggregated case is correctly approximating the original case in terms of generation scheduling and spot prices.

\begin{figure}[H]
\centering
\includegraphics[width=3.0in]{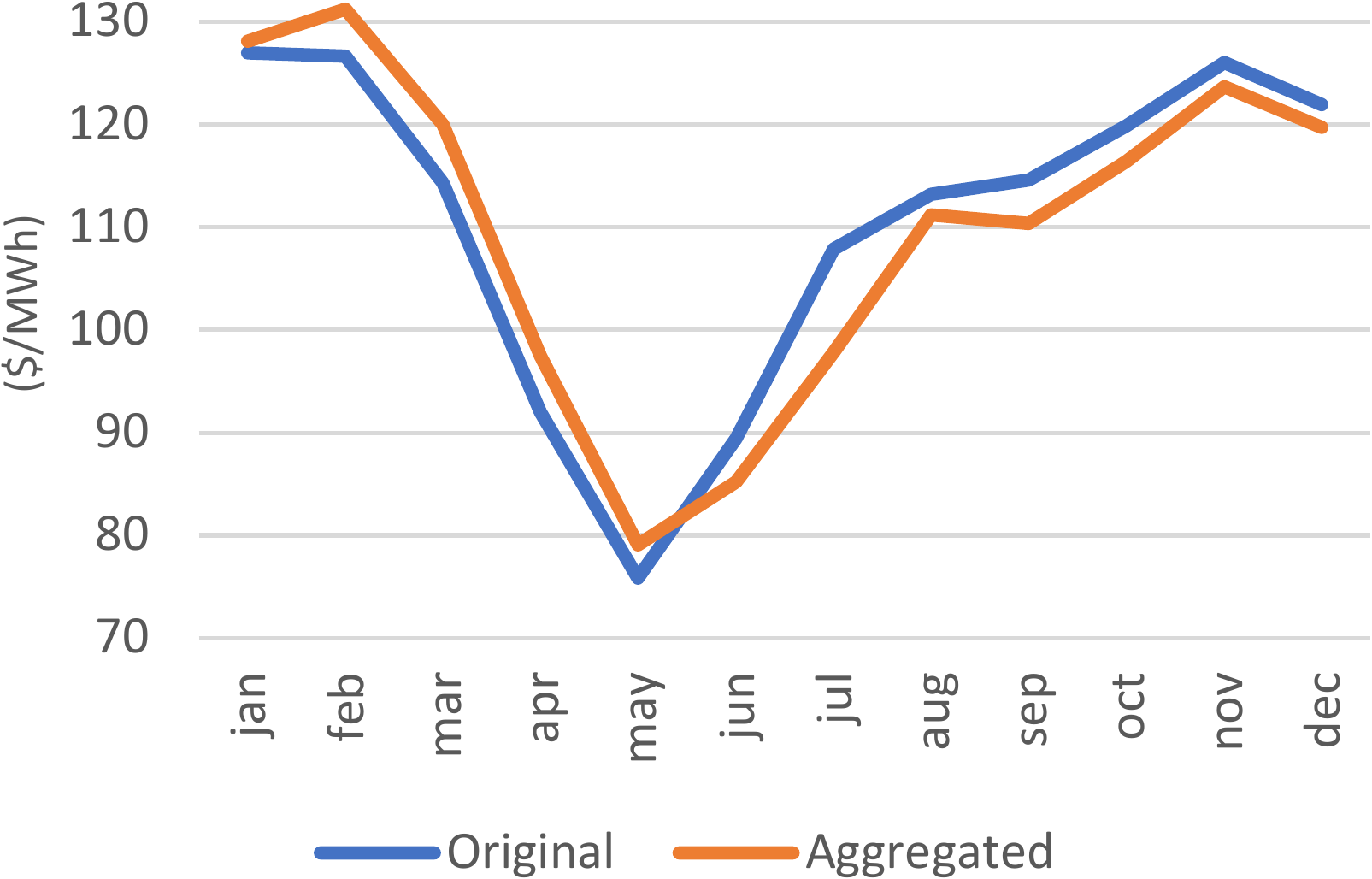}
\caption{Monthly marginal cost, in $\$/\text{MWh}$.}
\label{figure-marginal-cost}
\end{figure}

\tableref{figure-marginal-cost} compares the monthly marginal cost between the studies, in $\$/\text{MWh}$. The most considerable month is July, with a $9.35\%$ difference. In the remaining years, the maximum difference is $5.95\%$. 

\begin{figure}[H]
\centering
\includegraphics[width=3.0in]{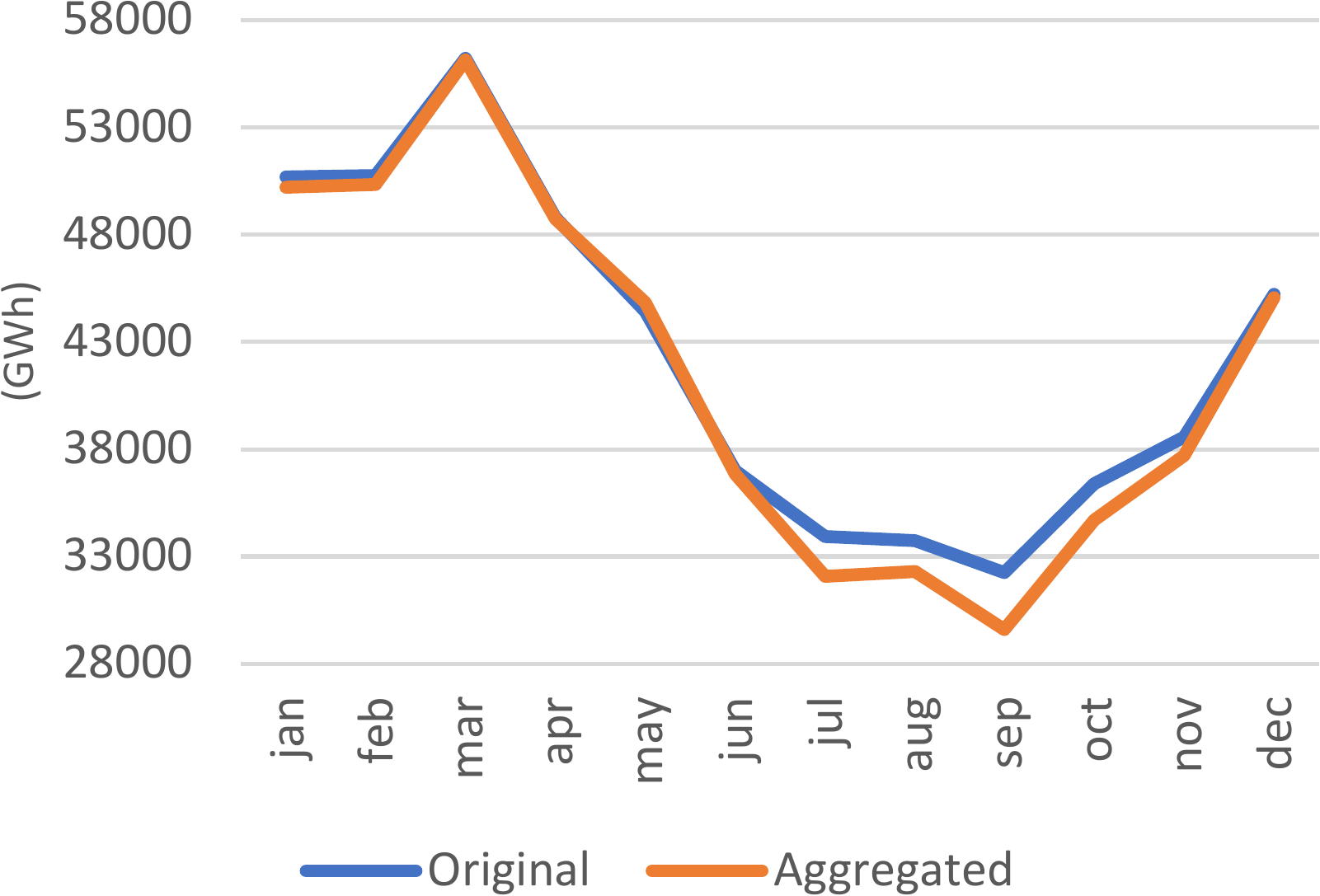}
\caption{Average hydroelectric generation, in GWh.}
\label{figure-hydro-generation}
\end{figure}

\begin{figure}[H]
\centering
\includegraphics[width=3.0in]{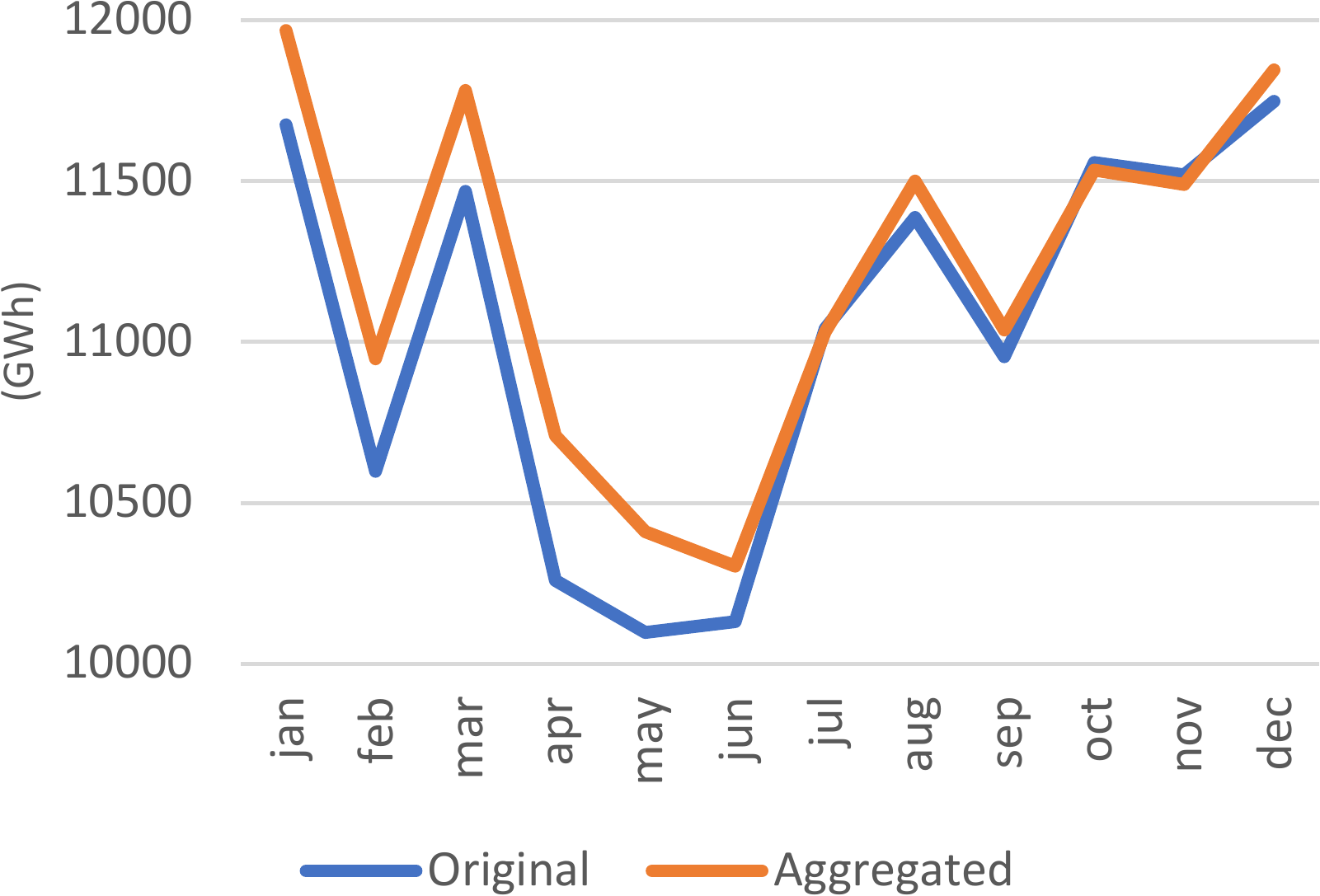}
\caption{Average thermal generation, in GWh.}
\label{figure-thermal-generation}
\end{figure}

\begin{figure}[H]
\centering
\includegraphics[width=3in]{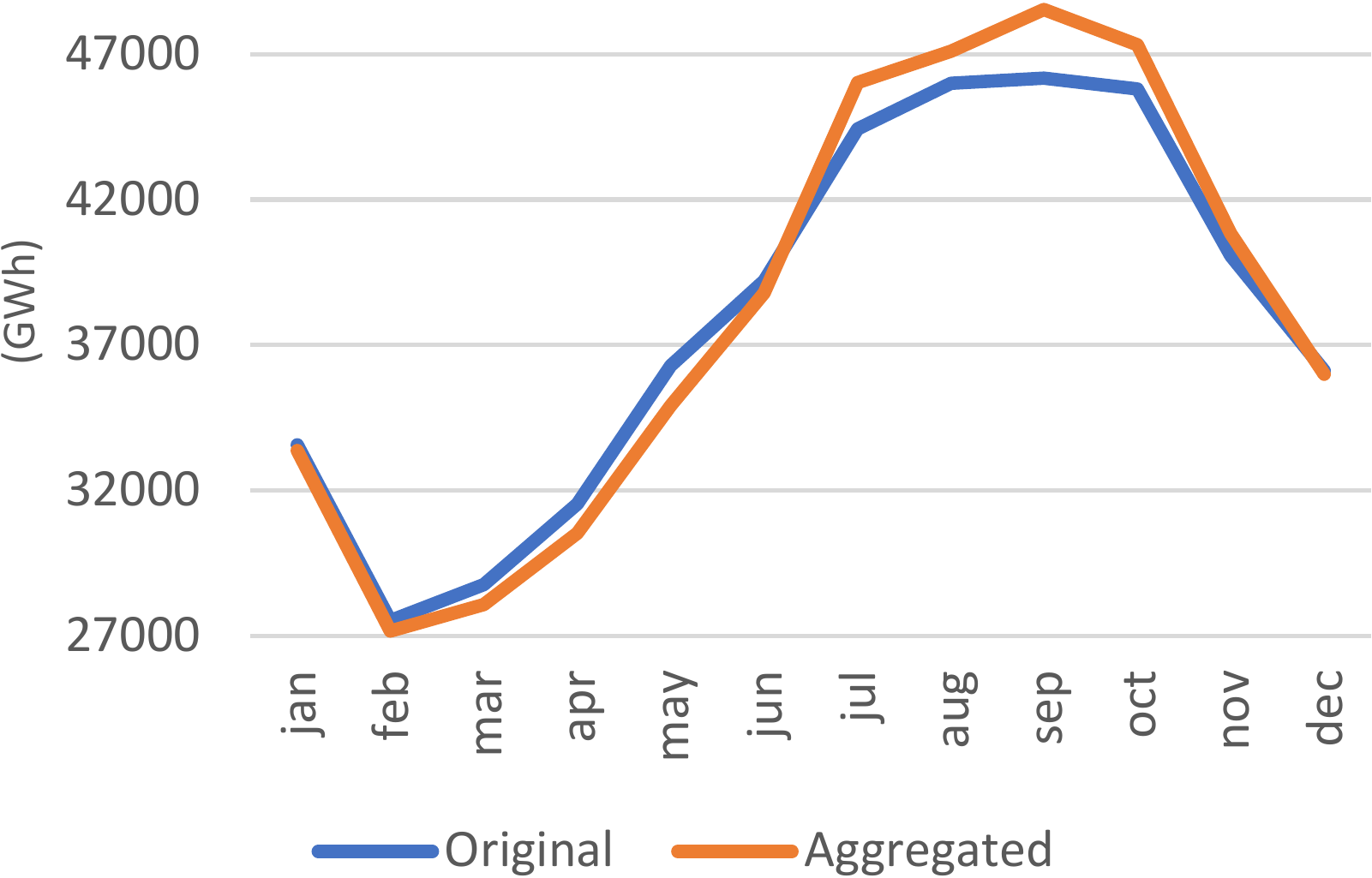}
\caption{Average renewable generation, in GWh.}
\label{figure-renew-generation}
\end{figure}

\figureref{figure-hydro-generation}, \figureref{figure-thermal-generation}, and \figureref{figure-renew-generation} displays the average hydroelectric, thermal, and renewable generation in GWh, respectively, of both studies. These figures indicate good approximation of the average generation.

The above-summarized results present that the proposed aggregation approach has established to be reasonably accurate in the Brazilian case study.
\section{Conclusion}
\label{section-conclusion}

We proposed a methodology to simplify network representations based on community detection and pre-computed spot-prices. The method was the applied to a detailed representation of the Brazilian power system.

The community detection algorithm was executed in reasonable time and produced a bus clustering that can be used in a power system simulation.

The simplified network was contrasted with the detailed one on a simulation of the power system operation for a large number of scenarios. The objective function of the underlying optimization problems, the system operative cost, was very well approximated varying less than one percent.

The simulation time of the aggregated problem was, as expected, much smaller than the complete system simulation. This indicates that the methodology can be used in simplified simulations where the execution time of a complete system is prohibitive.

The average behavior of spot prices and generation from different technologies show that the network simplification can be used to approximate many simulation outputs other than total cost.

Future works include applying the proposed method in other power systems, testing other variations of weights for the edges in the community detection algorithm and developing new methods to determine the capacity of interconnection between communities.

\section*{Acknowledgment}
The authors would like to thank the colleagues at PSR for the support and  conversations that led to this work.


\bibliographystyle{IEEEtran}
\bibliography{bibliography.bib}

\begin{thebibliography}{10}
\providecommand{\url}[1]{#1}
\csname url@samestyle\endcsname
\providecommand{\newblock}{\relax}
\providecommand{\bibinfo}[2]{#2}
\providecommand{\BIBentrySTDinterwordspacing}{\spaceskip=0pt\relax}
\providecommand{\BIBentryALTinterwordstretchfactor}{4}
\providecommand{\BIBentryALTinterwordspacing}{\spaceskip=\fontdimen2\font plus
\BIBentryALTinterwordstretchfactor\fontdimen3\font minus
  \fontdimen4\font\relax}
\providecommand{\BIBforeignlanguage}[2]{{%
\expandafter\ifx\csname l@#1\endcsname\relax
\typeout{** WARNING: IEEEtran.bst: No hyphenation pattern has been}%
\typeout{** loaded for the language `#1'. Using the pattern for}%
\typeout{** the default language instead.}%
\else
\language=\csname l@#1\endcsname
\fi
#2}}
\providecommand{\BIBdecl}{\relax}
\BIBdecl

\bibitem{gorenstin1993power}
B.~Gorenstin, N.~Campodonico, J.~Costa, and M.~Pereira, ``Power system
  expansion planning under uncertainty,'' \emph{IEEE Transactions on Power
  Systems}, vol.~8, no.~1, pp. 129--136, 1993.

\bibitem{soares2019addressing}
A.~Soares, R.~Perez, W.~Morais, and S.~Binato, ``Addressing the time-varying
  dynamic probabilistic reserve sizing method on generation and transmission
  investment planning decisions,'' \emph{arXiv preprint arXiv:1910.00454},
  2019.

\bibitem{cheng2005ptdf}
X.~Cheng and T.~J. Overbye, ``Ptdf-based power system equivalents,'' \emph{IEEE
  Transactions on Power Systems}, vol.~20, no.~4, pp. 1868--1876, 2005.

\bibitem{lloyd1982least}
S.~Lloyd, ``Least squares quantization in pcm,'' \emph{IEEE transactions on
  information theory}, vol.~28, no.~2, pp. 129--137, 1982.

\bibitem{stockertransmission}
N.~Stocker, ``Transmission network model simplification and approximation,''
  Master's thesis, Swiss Federal Institute of Technology (ETH), Zurich, 2016.

\bibitem{cotilla2013multi}
E.~Cotilla-Sanchez, P.~D. Hines, C.~Barrows, S.~Blumsack, and M.~Patel,
  ``Multi-attribute partitioning of power networks based on electrical
  distance,'' \emph{IEEE Transactions on Power Systems}, vol.~28, no.~4, pp.
  4979--4987, 2013.

\bibitem{cao2018incorporating}
K.-K. Cao, J.~Metzdorf, and S.~Birbalta, ``Incorporating power transmission
  bottlenecks into aggregated energy system models,'' \emph{Sustainability},
  vol.~10, no.~6, p. 1916, 2018.

\bibitem{entsoe2018}
{European Network of Transmission System Operators for Electricity}, ``First
  edition of the bidding zone review,''
  \url{https://www.entsoe.eu/news/2018/04/05/first-edition-of-the-bidding-zone-review-published/},
  2018, accessed: 2019-09-27.

\bibitem{prim1957shortest}
R.~C. Prim, ``Shortest connection networks and some generalizations,''
  \emph{The Bell System Technical Journal}, vol.~36, no.~6, pp. 1389--1401,
  1957.

\bibitem{newman2004finding}
M.~E. Newman and M.~Girvan, ``Finding and evaluating community structure in
  networks,'' \emph{Physical review E}, vol.~69, no.~2, p. 026113, 2004.

\bibitem{anthonisse1971rush}
J.~M. Anthonisse, ``The rush in a directed graph,'' \emph{Stichting
  Mathematisch Centrum. Mathematische Besliskunde}, no. BN 9/71, 1971.

\bibitem{freeman1977set}
L.~C. Freeman, ``A set of measures of centrality based on betweenness,''
  \emph{Sociometry}, pp. 35--41, 1977.

\bibitem{girvan2002community}
M.~Girvan and M.~E. Newman, ``Community structure in social and biological
  networks,'' \emph{Proceedings of the national academy of sciences}, vol.~99,
  no.~12, pp. 7821--7826, 2002.

\bibitem{brandes2001faster}
U.~Brandes, ``A faster algorithm for betweenness centrality,'' \emph{Journal of
  mathematical sociology}, vol.~25, no.~2, pp. 163--177, 2001.

\bibitem{fortunato2010community}
S.~Fortunato, ``Community detection in graphs,'' \emph{Physics reports}, vol.
  486, no. 3-5, pp. 75--174, 2010.

\bibitem{pereira1991multi}
M.~V. Pereira and L.~M. Pinto, ``Multi-stage stochastic optimization applied to
  energy planning,'' \emph{Mathematical programming}, vol.~52, no. 1-3, pp.
  359--375, 1991.

\end{thebibliography}

\end{document}